\documentclass[fleqn,10pt]{wlscirep}
\usepackage[utf8]{inputenc}
\usepackage[T1]{fontenc}
\usepackage{multirow}
\usepackage{adjustbox}

\makeatletter
\newcommand*{\rom}[1]{\expandafter\@slowromancap\romannumeral #1@}
\makeatother

\title{Raidionics: an open software for pre- and postoperative central nervous system tumor segmentation and standardized reporting}

\author[1]{David Bouget}
\author[1]{Demah Alsinan}
\author[1]{Valeria Gaitan}
\author[1,2]{Ragnhild Holden Helland}
\author[1]{André Pedersen}
\author[3,4]{Ole Solheim}
\author[1,2,*]{Ingerid Reinertsen}
\affil[1]{SINTEF Digital, Department of Health Research, NO-7465 Trondheim, Norway}
\affil[2]{Norwegian University of Science and Technology (NTNU), Department of Circulation and Medical Imaging, NO-7491 Trondheim, Norway}
\affil[3]{St. Olavs hospital, Trondheim University Hospital, Department of Neurosurgery, NO-7491 Trondheim, Norway}
\affil[4]{Norwegian University of Science and Technology (NTNU), Department of Neuromedicine and Movement Science, NO-7491 Trondheim, Norway}

\affil[*]{ingerid.reinertsen@ntnu.no}


\begin{abstract}
For patients suffering from central nervous system tumors, prognosis estimation, treatment decisions, and postoperative assessments are made from the analysis of a set of magnetic resonance (MR) scans. Currently, the lack of open tools for standardized and automatic tumor segmentation and generation of clinical reports, incorporating relevant tumor characteristics, leads to potential risks from inherent decisions' subjectivity.
To tackle this problem, the proposed Raidionics open-source software has been developed, offering both a user-friendly graphical user interface and stable processing backend. The software includes preoperative segmentation models for each of the most common tumor types (i.e., glioblastomas, lower grade gliomas, meningiomas, and metastases), together with one early postoperative glioblastoma segmentation model. 
Preoperative segmentation performances were quite homogeneous across the four different brain tumor types, with an average Dice around 85\% and patient-wise recall and precision around 95\%. Postoperatively, performances were lower with an average Dice of 41\%. Overall, the generation of a standardized clinical report, including the tumor segmentation and features computation, requires about ten minutes on a regular laptop.
The proposed Raidionics software is the first open solution enabling an easy use of state-of-the-art segmentation models for all major tumor types, including preoperative and postsurgical standardized reports.
\end{abstract}
\begin{document}

\flushbottom
\maketitle
%
%
\thispagestyle{empty}



\section*{Introduction}
Central nervous system (CNS) tumors, further classified into taxonomic categories as per iterative editions from the World Health Organization~\cite{louis20212021}, depict all possible tumors originating from the brain or spinal cord. Given more than 100 subtypes, the heterogeneity in a tumor expression (i.e., location, growth rate, or invasiveness) leads to a likewise heterogeneous prognosis. Most patients will experience neurological and cognitive deficits over time~\cite{day2016neurocognitive}, with survival expectancy ranging from weeks to several years depending on the tumor type and grade. Primary tumors, emanating from the brain itself or its supporting tissues, represent the vast majority of CNS tumors. As opposed to secondary tumors, arising elsewhere in the body and then transferred to the brain (i.e., metastases). In the former, the most frequent subtypes arise either from the brain's glial tissue (i.e., gliomas) or from the meninges (i.e., meningiomas). The most aggressive gliomas, further categorized as glioblastomas (noted GBM), remain among the most difficult cancers to treat with an extremely short overall survival~\cite{lapointe2018primary}. Less aggressive entities, categorized as diffuse lower-grade gliomas (noted LGG), are infiltrative neoplasms like other gliomas, highly invasive, and impossible to resect~\cite{cancer2015comprehensive}.
Initial tumor discovery, treatment decisions, and preoperative prognosis assessment are based on the analysis of a set of magnetic resonance (MR) scans. For maximizing patient outcome and facilitating optimal treatment decisions, the utmost accuracy during the diagnostics phase is imperative from the multidisciplinary team of surgeons, radiologists, and oncologists. The coupling of MR scans to genetic and histopathological findings from tissue analysis has shown benefits to narrow the tumor classification and presence of mutations, further assisting to refine clinical outcomes and guide clinical decision making~\cite{appin2014molecular,jiao2012frequent}.
Currently, informative tumor characteristics are estimated from the MR scans either through crude measuring techniques (e.g., eyeballing or short-axis diameter estimation) or after manual tumor delineation. However, such procedures are either inherently time-consuming or often liable to intra and inter-rater variability. A lack of user-friendly software solutions for retrieving quantitative and standardized information for patients with intracranial tumors stands out as a major hurdle preventing widespread access in clinical practice, clinical research, or over tumor registries.

The task of automatic brain tumor segmentation from preoperative MR scans is an actively researched field~\cite{wadhwa2019review, tiwari2020brain, havaei2017brain}. Multiple previous studies did not disambiguate between CNS tumor types~\cite{pereira2016brain,myronenko20193d,ranjbarzadeh2021brain,naser2020brain,sun2019brain} and trained generic segmentation models, investigating at the same time other tasks such as classification or survival estimation. Most studies investigating brain tumor segmentation globally have used the public dataset from the BraTS challenge~\cite{menze2014multimodal}. A constant attention is upheld by the community thanks to the MICCAI challenge occuring every year since 2012, promoting research on glioma sub-regions segmentation and classification to predict clinical biomarkers status. The dataset contains a patient cohort of up to 2040 patients and multiple MR sequences included for each patient: T1-weighted (T1w), gadolinium-enhanced T1-weighted (T1c), T2-weighted fluid attenuated inversion recovery (FLAIR), and T2-weighted (T2). As a result, the most studied CNS tumor in the literature is by far the glioma, including both GBM and LGG.
The current state-of-the-art baseline method for tumor segmentation is the nnU-Net framework~\cite{isensee2021nnu}, which is a typical encoder decoder architecture coupled to a smart parameters optimization scheme for preprocessing and training, to cater to the input dataset. Average Dice scores about 90\% have been reached over contrast-enhancing tumor, necrosis, or edema sub-regions.
For the meningioma subtype, a literature review has made an inventory of all studies performed between 2008 and 2020~\cite{neromyliotis2022machine}. At best, no more than 130 patients were included to train models using widespread 3D neural network architectures, achieving average Dice scores around 90\%~\cite{laukamp2019fully, laukamp2021automated}. In our recent study~\cite{bouget2021meningioma}, a much larger dataset with 700 patients was used, for similar overall performances. The validation was extended to show robustness across MR scan resolution and tumor volume. Finally, brain metastasis segmentation has been investigated over multicentric and multi-sequence datasets of up to 200 patients~\cite{charron2018automatic, liu2017deep, grovik2020deep, grovik2021handling}, achieving on average up to 80\% Dice score using either the DeepLabV3~\cite{chen2017rethinking} or the DeepMedic architecture~\cite{kamnitsas2016deepmedic}.

To summarize, the task of CNS tumor segmentation has been well investigated on preoperative MR scans, favoring GBM and LGG subtypes through unrestricted access to an open and annotated dataset. Conversely, segmentation in postoperative MR scans has been scarcely addressed as of yet due to its unparalleled difficulty and lack of public data. Recently, Lotan et al. proposed to fuse two of the top-ranked BraTS methods for performing both pre- and postoperative GBM segmentation~\cite{lotan2022development}. Over the 20 postoperative MR samples constituting the test set, an average Dice score of 74\% was reached over the contrast-enhancing subregion. Having access to a dataset of a larger magnitude including 500 patients, an average Dice score of 69\% was reported using all MR sequences as input and an ensemble of nnU-Net models~\cite{nalepa2023deep}. Finally, similar performances were reached on a dataset including up to 900 patients and multiple MR sequences, also leveraging the nnU-Net architecture~\cite{helland2023segmentation}.
Being able to generate high-quality automatic segmentations is a mandatory initial step to provide reproducible and trustworthy standardized reports to characterise the tumor (RADS). The ultimate objective is to assist the clinical team in making the best assessment regarding treatment options and patient outcome. However, the segmentation quality has only very recently reached an acceptable threshold and as such the literature on RADSs for CNS tumors is scarce. For preoperative glioblastoma surgery, tumor features such as volume, multifocality, and location with respect to cortical and subcortical structures was presented~\cite{kommers2021glioblastoma}. An excellent agreement between features collected from the automatic segmentation and the manual segmentation was reported. For post-treatment investigations, a structured set of rules was suggested, deprived of any automatic segmentation or image analysis support~\cite{weinberg2018management}.

For use in routine clinical practice, the aforementioned segmentation models or RADS methods must be packaged into well-rounded solutions directly usable by most practitioners. A web imaging platform for radiology is being developed, Open Health Imaging Foundation (OHIF)~\cite{urban2017lesiontracker}, leaving the possibility to interface developed methods through custom plugins, and run processes on a server either locally to a hospital with access to PACS or remote. MONAI, a multipotent toolbox covering a wide range of use-cases including brain tumors, is being actively developed~\cite{cardoso2022monai}. While the MONAI Label component is meant as a development tool for creating or refining segmentation models through manual annotation, the MONAI Deploy component focuses on bringing AI-driven applications into the healthcare imaging domain. Even though custom plugins can be developed in both solutions, no trained models for CNS tumor segmentation or standardized reporting are available.
On the other hand, less advanced or refined solutions have been developed, focusing on the task at hand. A toolkit has been developed for running preoperative GBM segmentation models from the BraTS challenge, with a very minimalistic graphical user interface (GUI)~\cite{kofler2020brats}.
Inside the 3D Slicer software~\cite{pieper20043d}, often used by clinicians for performing semi-manual tumor delineation, plugins have been developed to facilitate the deployment of custom models with DeepInfer~\cite{mehrtash2017deepinfer} or with existing brain tumor segmentation models with DeepSeg~\cite{zeineldin2021slicer}.
To conclude, some focus has recently been dedicated to the accessibility of developed methods, yet not many solutions are providing trained segmentation models for CNS tumors other than preoperative gliomas.
In instances where trained segmentation models and inference scripts are publicly available, some extent of computer science and programming knowledge is required for running inference scripts locally on MR scans. However, this process is too overwhelming for most clinicians and hospital practitioners. Finally, no open solution exists offering the possibility to perform clinical reporting in a standardized fashion (i.e., RADS).

Upon initial publication~\cite{bouget2022preoperative}, a Raidionics prototype was introduced, first open-source solution, offering the possibility to segment the most frequent brain tumor types in preoperative MR scans (namely glioblastoma, diffuse lower-grade glioma, meningioma, and metastasis). One RADS mode was also available for describing the segmented tumor in terms of overall location in the brain and respective location against cortical and subcortical structures.
In this paper, the first complete and stable Raidionics software version is presented, including the following novelties. First, (i) the GUI has been completely redesigned, requiring only a few clicks and no programming skills to run segmentation and reporting tasks, either for single patients or entire cohorts. In the meantime, the processing backbone remains independently available to users with programming experience or for PACS integration. Second, (ii) the preoperative segmentation models have been improved, trained and validated using various datasets, reaching performances on-par or better than state-of-the-art reported performances. Third, (iii) an early postoperative glioblastoma residual tumor segmentation model has been included; the first open-access model for the task. Finally, (iv) a standardized report for postsurgical assessment has been incorporated.

\section*{Data}
In this work, four curated datasets were leveraged to train and develop the proposed methods, one for each considered CNS tumor type.
For glioblastoma (GBM), 2125 T1c patient scans were gathered from 15 institutions. For diffuse lower-grade glioma (LGG), 678 FLAIR patient scans were compiled from four institutions. For meningioma, 706 T1c patient scans were retrieved from two institutions, and finally 394 T1c patient scans were collected from two institutions for metastasis.

All tumors were manually delineated by trained raters, under supervision of neuroradiologists and neurosurgeons. The tumor was defined as gadolinum-enhancing tissue, including non-enhancing enclosed necrosis or cysts in T1c scan and as the hyperintense region in FLAIR scan. Initial segmentations were performed using either a region growing algorithm~\cite{huber2017reliability} or a grow-cut algorithm~\cite{vezhnevets2005growcut}, followed by manual correction.

\begin{figure}[!t]
\centering
\includegraphics[scale=0.465]{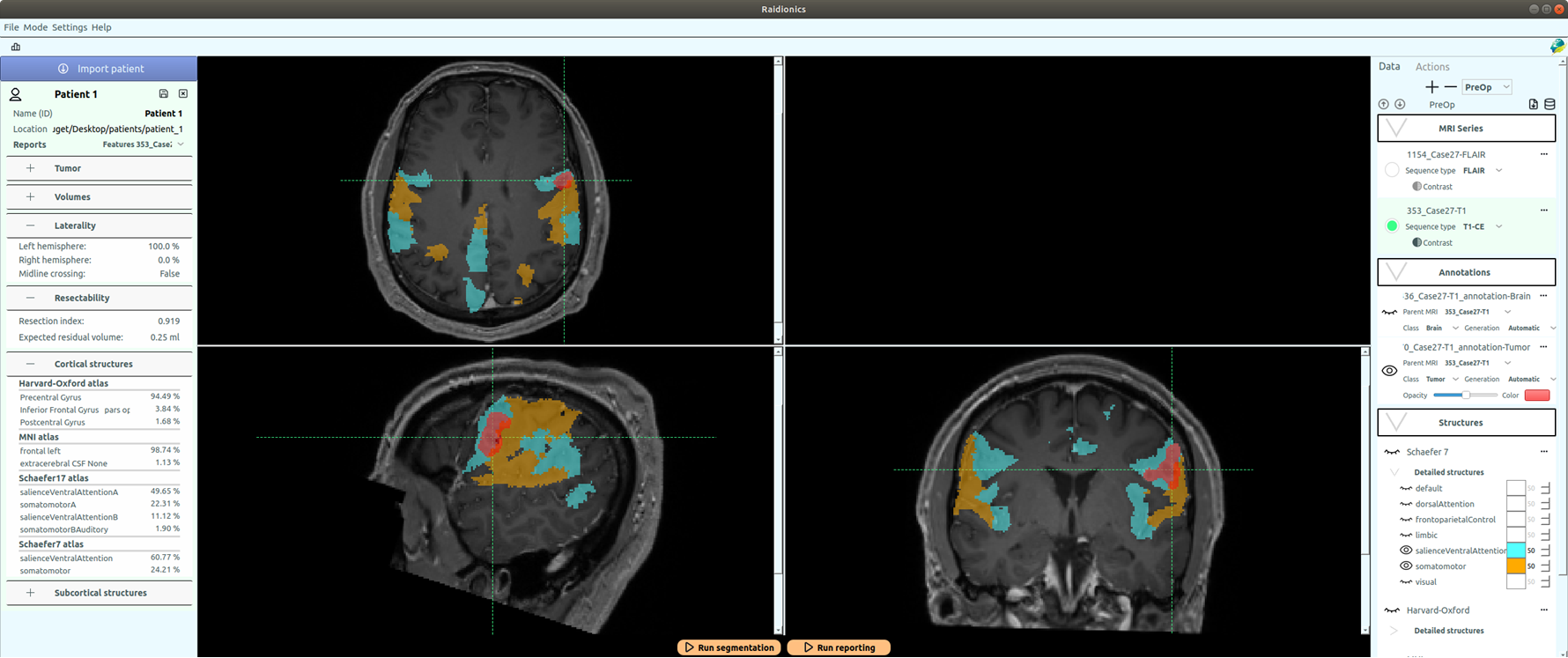}
\caption{Illustration of the Raidionics software GUI in single patient mode, after generating the standardized report over a glioblastoma case. The left side presents the tumor characteristics belonging to the report, the central part offers a set of 2D views, and the right panel shows loaded MR scans and corresponding annotations.}
\label{fig:raidionics-illu}
\end{figure}

\section*{Methods}
To make the segmentation models and standardized generation of clinical reports easily available to a wider audience, the Raidionics software has been developed with a special care towards the user interface design. In the following sections, the different components of the software are described, the strategy for training the segmentation models is presented, and both pipelines for pre- and postoperative clinical reporting are explained. 

\subsection*{Modes}
\begin{figure}[!t]
\centering
\includegraphics[scale=0.35]{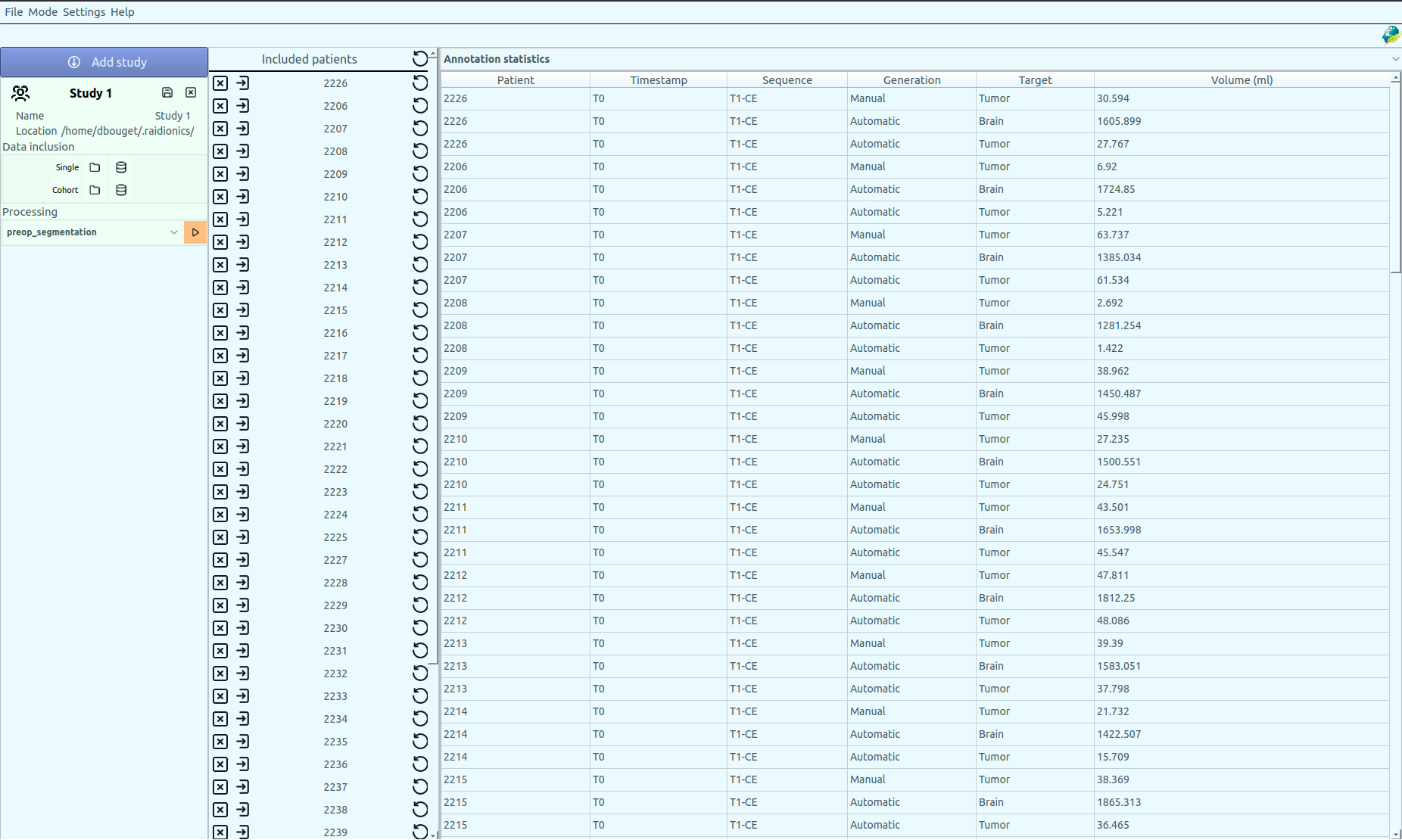}
\caption{Illustration of the Raidionics software GUI in batch/study mode, after processing a cohort of glioblastoma patients. The left side gathers the study options, the central part lists the patients currently included in the study, and the right panel offers a summary of the processing results.}
\label{fig:raidionics-illu-batchmode}
\end{figure}

In Raidionics, two main modes are available: single patient mode (illustrated in Fig.~\ref{fig:raidionics-illu}) and batch/study mode (illustrated in Fig.~\ref{fig:raidionics-illu-batchmode}).
In single patient mode, direct visualization and interaction with patient's data and corresponding results is available. In this mode, the GUI is split into three main components, starting with the left side panel relating to patient import and browsing of standardized reports. Patients can be saved, closed, reloaded, and renamed for only one patient displayed at any given time. In the center panel, three 2D viewers are proposed for displaying the selected MR scan following standard axial, coronal, and sagittal slicing. All views are interconnected and aligned under the same 3D location, adjustable by mouse clicking, and represented by the two cross-hair green dotted lines.
Finally, the panel on the right side of the interface lists all MR scans and corresponding annotations or structural atlases for each scan, for the given patient. While only one MR scan can be toggled visible at the time, multiple annotations or structural atlases can be freely overlaid. Each overlaid item can be customized in color and opacity, improving the display, and allowing the generation of illustrations. By selecting the Actions tab, automatic segmentation or standardized reporting processes can be launched for the current patient.
In batch/study mode, cohorts of patients can be loaded and processed sequentially, without any direct visualization or possibility to interact with the results. The GUI is likewise split into three components in this mode with a left side panel relating to study creation and import. For each study, patients can be imported either by careful selection of a few patient folders, or by selection of a cohort folder (i.e., containing multiple folders, one per patient). The launching of a segmentation or reporting process over an entire cohort can also be performed from within each study. In the center panel, all included patients for the current study are listed, can be removed, and can be opened in single patient mode for viewing and interaction purposes. Finally, the right panel proposes different summary tables, starting with a content summary listing all files included for all patients. In addition, annotation and standardized reporting tables are included to provide an overview of all extracted parameters for the patients of a given cohort.

\subsection*{Data import}
Data can be imported in Raidionics in different formats, either as raw DICOM folders originating from PACS or as converted volumes in popular formats such as NIfTI (.nii, .nii.gz), MetaImage (.mhd), or NRRD (.nrrd). More generally, all formats accepted in SimpleITK~\cite{yaniv2018simpleitk}, underlying Python library used for reading converted volumes, are likewise approved. Upon import, all MR scans are internally converted to NIfTI format for subsequent processing. Annotation files can also be imported manually, requiring matching volume parameters (i.e., shape, spacing, and orientation) to the corresponding MR scan. Upon loading, if multiple MR scans were imported for the current patient, the annotation must be linked to the proper MR scan using the drop-down selector (named Parent MRI, in the right side panel).

Within Raidionics, all MR scans are expected to be associated to data timestamps organized in ascending order, allowing the disambiguation between preoperative and postoperative content loaded simultaneously for a patient. Timestamps can be manually edited in the right side panel of the single patient mode interface.

Overall, four approaches are available for loading patient data into the software. First, a DICOM folder can be selected, and a pop-up window will allow for selecting specific MR acquisitions or importing all available acquisitions in a bulk. Second, converted MR scans can be individually selected, automatically linked to the current data timestamp. Third, an entire folder can be selected, either containing multiple converted MR scan files or multiple sub-folders. In the latter, the data is assumed to be split into data timestamps, and loaded data will be organized as such. Finally, reopening a patient folder, previously saved through Raidionics, can be done by selecting the corresponding custom scene file (.raidionics).

\begin{figure}[!b]
\centering
\includegraphics[scale=0.55]{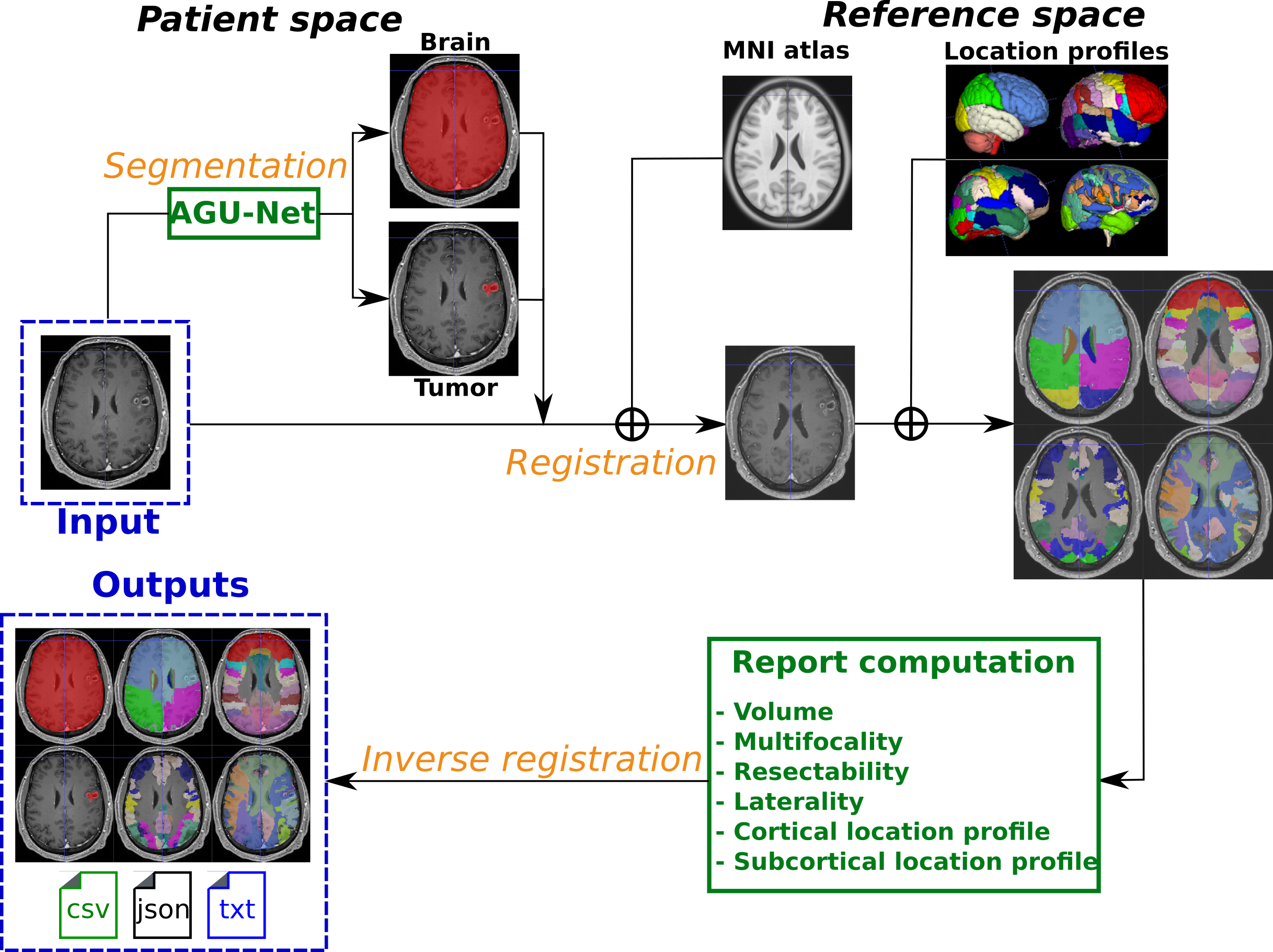}
\caption{Overall segmentation and standardized report generation pipeline. The AGU-Net architecture was used for the segmentation task whereas the image registration task was performed using the SyN method from ANTs.}
\label{fig:pipeline-illu}
\end{figure}

\subsection*{Storing results}
For each patient, all results are stored inside the corresponding folder, including a custom .raidionics file for fast reloading. This concept is similar to scene files from 3D Slicer stored as .mrml on disk. All volume files are stored as NIfTI (i.e., MR scan, segmentations, and atlases), statistics are stored as comma-separated values files (.csv), and standardized reports are stored as text files (.txt), json files (.json), and csv files (.csv).

\subsection*{Tumor segmentation}
All preoperative CNS tumor segmentation models included in the software have been trained with the AGU-Net architecture~\cite{bouget2021meningioma} using five levels with $\{16, 32, 128, 256, 256\}$ as filter sizes, deep supervision, multiscale input, and single attention modules. Unlike the originally published architecture, all batch normalization layers have been removed and a patch-wise approach for training and inference was followed, with $160^{3}$\,voxels as patch dimension. The preprocessing was limited to a $0.75\,\text{mm}^{3}$ isotropic resampling, skull-stripping (except for the meningioma subtype), intensity clipping to remove the 0.05\% highest values, and finally intensity normalization and scaling to [0, 1].
Training has been performed with batch size four and using the Adam optimizer with an initial learning rate of $5 \cdot 10^{-4}$. In addition, a gradient accumulation of 8 steps was performed, resulting in a virtual batch size of 32 samples, using the open TensorFlow model wrapper implementation~\cite{andre_pedersen_2023_7831244}. The number of updates per epoch has been limited to 512, and an early stopping scheme was setup to stop training after 15 consecutive epochs without validation loss improvement.

For early postoperative segmentation of glioblastomas, the model available in Raidionics has been introduced in a recent study~\cite{helland2023segmentation}.

\subsection*{Features extraction and standardized reporting (RADS)}
The overall process for segmentation and standardized report generation with relevant tumor characteristics is depicted in Fig.~\ref{fig:pipeline-illu}.
For the generation of standardized preoperative clinical reports in a reproducible fashion, the computation of tumor characteristics was performed after alignment to a standard reference space, the symmetric Montreal Neurological Institute (MNI) ICBM2009a atlas~\cite{fonov2009unbiased,fonov2011unbiased}. The patient's input MR scan was registered to the corresponding atlas file using the SyN method from the Advanced Normalization Tools (ANTs)~\cite{avants2008symmetric}. The collection of computed tumor features includes: volume, laterality, multifocality, cortical structure location profile, and subcortical structure location profile. Specifically tailored for glioblastomas, resectability features are therefore not available for the other CNS tumor types. A more in-depth description of the computed parameters is available in our previous study~\cite{bouget2021glioblastoma}.

For postsurgical assessment, both preoperative and postoperative volumes, extent of resection (EOR), and EOR patient classification are automatically extracted, following the latest guidelines~\cite{karschnia2022surg}.

\subsection*{Deployment}
Installation executables have been created for cross-platform use of Raidionics, compatible with Windows ($\geq$ 10), Ubuntu Linux ($\geq$ 18.04), and macOS ($\geq$ Catalina 10.15) including ARM-based Apple M1. The selected inference engine to run the segmentation models is ONNX Runtime, supporting models from various deep learning frameworks and widely compatible across hardware, drivers, and operating systems.

The core computational backend (i.e., without any GUI) is also available for experienced users, allowing for direct use either through the command-line interface, as a Python library, or as a Docker container. In addition, a Raidionics 3D Slicer plugin is available, directly leveraging the backend Docker container.

\section*{Results}
Experiments were carried out on multiple machines with the following specifications: Intel Xeon W-2135 CPU @3.70 GHz x 12, 64 GB of RAM, NVIDIA Tesla V100S (32GB), and a regular hard-drive.
The implementation was done in Python 3.7, using PySide6 v6.2.4 for the GUI, TensorFlow v2.8 for training the segmentation models, and ONNX Runtime v1.12.1 for running model inference.

\subsection*{Segmentation performance}
All preoperative CNS segmentation models were trained from scratch under the same k-fold cross-validation paradigm whereby one fold was used as validation set, one fold as test set, and all remaining folds constituted the training set. For the glioblastoma subtype, a leave-one-hospital-out cross-validation paradigm was followed, equivalent to a 15-fold cross-validation.
Pooled estimates, computed from each fold's results, are reported for each measurement~\cite{killeen2005alternative}. Overall, measurements are reported as mean and standard deviation (indicated by $\pm$) in the tables.

A summary of the segmentation performances for the models packaged in Raidionics is provided in Table~\ref{tab:all-seg-results}. For all preoperative models, an average Dice score of 85\% and patient-wise F1-score of 95\% are achieved, highlighting a high segmentation quality. The lowest Dice score of 78\% is obtained for the LGG subtype, which can be explained by the diffuse nature of such tumors, more difficult to fully delineate in FLAIR MR scans. Overall, segmentation performances are largely stable across the different CNS tumor subtypes, with extremely accurate sensitivity and specificity of the different models. The models are quite conservative with few false positives, and simultaneously efficient with few tumors missed. For multifocal tumors, often with satellite foci clearly smaller than the main focus, an average object-wise recall of 80\% is achieved, indicating a struggle to properly segment tiny structures. A similar decrease in object-wise precision can be acknowledged, around 87\% on average, symptomatic of a segmentation excess over false positive areas, locally resembling contrast-enhancing tumor tissue.

\begin{table}[!ht]
\caption{Overall segmentation performance summary for each CNS tumor type. The bottom line reports performances obtained with the published postoperative GBM segmentation model~\cite{helland2023segmentation}.}
\adjustbox{max width=\textwidth}{
\begin{tabular}{l|cc|ccc|ccc}
 & \multicolumn{2}{c|}{Voxel-wise} & \multicolumn{3}{c|}{Patient-wise} & \multicolumn{3}{c}{Object-wise} \tabularnewline
Tumor type & Dice & Dice-TP & F1-score & Recall & Precision & F1-score & Recall & Precision\tabularnewline
\hline
Meningioma & $82.18\pm23.67$ & $86.62\pm14.54$ & $93.51\pm02.83$ & $94.90\pm03.29$ & $92.32\pm04.32$ & $86.04\pm04.50$ & $88.36\pm03.33$ & $84.32\pm07.99$\tabularnewline
Metastasis & $86.55\pm19.10$ & $88.58\pm14.08$ & $96.60\pm01.66$ & $97.73\pm02.09$ & $95.61\pm03.46$ & $87.47\pm04.88$ & $82.94\pm05.13$ & $92.67\pm05.58$ \tabularnewline
LGG & $78.71\pm21.40$ & $81.61\pm15.54$ & $94.60\pm01.37$ & $96.46\pm02.29$ & $92.88\pm02.26$ & $82.41\pm06.33$ & $80.93\pm09.96$ & $84.77\pm05.59$ \tabularnewline
GBM preop. & $85.17\pm16.58$ & $86.63\pm12.41$ & $96.79\pm01.27$ & $98.31\pm01.10$ & $95.35\pm02.13$ & $88.25\pm05.09$ & $85.06\pm07.87$ & $91.96\pm03.33$\tabularnewline
\hline
GBM postop. & $41.02\pm28.08$ & $52.45\pm20.14$ & $83.73\pm03.17$ & $82.80\pm05.27$ & $85.16\pm05.24$ & --- & --- & --- \tabularnewline
\end{tabular}
}
\label{tab:all-seg-results}
\end{table}

To further investigate segmentation performances from a tumor volume standpoint, an empirical categorization was made to single out small tumors. The volume cut-off was set to 2\,ml for all CNS tumor subtypes, except for the LGG subtype where it was set to 5\,ml to feature enough cases in the small tumor category. The categorized segmentation performances based on tumor volume are reported in Table~\ref{tab:categorical-seg-results}. Unsurprisingly, segmentation performances obtained over non-small tumors are excellent with 99\% patient-wise recall and up to 90\% Dice-TP for the metastasis subtype. In comparison, the average Dice score for the small tumors category lies closer to 60\%, achieving barely 75\% patient-wise recall globally. Such performances are less enticing as they highlight limitations for using the packaged models to perform early-stage tumor detection. However, larger performance discrepancies for the small tumors category across the different CNS subtypes can be observed. The lowest Dice and patient-wise recall values are repeatedly obtained for the LGG subtype, whereas a 92\% patient-wise recall is obtained for the metastasis subtype.
Compared to previous publications using the AGU-Net architecture in a full volume fashion over the same task~\cite{bouget2021meningioma,bouget2022preoperative}, using a patch-wise strategy improved segmentation performances overall, especially over small tumors.

\begin{table}[!ht]
\caption{Preoperative segmentation performances summary for each CNS tumor subtype based on two tumor volume categories. The number of patients inside each category is indicated in parenthesis.}
\adjustbox{max width=\textwidth}{
\begin{tabular}{ll|cc|ccc|ccc}
 & & \multicolumn{2}{c|}{Voxel-wise} & \multicolumn{3}{c|}{Patient-wise} & \multicolumn{3}{c}{Object-wise} \tabularnewline
Type & Volume & Dice & Dice-TP & F1-score & Recall & Precision & F1-score & Recall & Precision\tabularnewline
\hline
\multirow{2}{*}{Meningioma} & <2\,ml (139) & $63.23\pm33.65$ & $77.86\pm17.91$ & $85.55\pm08.96$ & $81.29\pm12.86$ & $91.52\pm08.27$ & $81.29\pm08.34$ & $77.45\pm11.07$ & $87.06\pm11.21$ \tabularnewline & $\geq$2\,ml (567) & $86.82\pm17.26$ & $88.38\pm12.94$ & $95.21\pm02.53$ & $98.23\pm01.53$ & $92.46\pm04.20$ & $86.94\pm04.96$ & $90.70\pm03.61$ & $83.96\pm08.45$ \tabularnewline
\hline
\multirow{2}{*}{Metastasis} & <2\,ml (50) & $70.07\pm30.89$ & $76.69\pm24.93$ & $93.92\pm06.35$ & $92.00\pm09.58$ & $96.71\pm07.73$ & $80.30\pm13.37$ & $72.65\pm21.54$ & $95.38\pm10.40$\tabularnewline & $\geq$2\,ml (344) & $89.46\pm13.20$ & $90.26\pm10.30$ & $97.20\pm01.76$ & $99.12\pm1.26$ & $95.46\pm03.71$ & $88.73\pm05.08$ & $85.35\pm05.31$ & $92.55\pm06.07$\tabularnewline
\hline
\multirow{2}{*}{LGG} & <5\,ml (74) & $50.58\pm32.95$ & $64.40\pm23.84$ & $82.78\pm17.35$ & $77.02\pm21.82$ & $94.11\pm12.20$ & $79.16\pm17.81$ & $73.24\pm21.02$ & $91.50\pm16.18$\tabularnewline & $\geq$5\,ml (604) & $82.15\pm16.24$ & $83.11\pm13.72$ & $95.66\pm01.34$ & $98.84\pm01.07$ & $92.72\pm02.32$ & $82.70\pm07.66$ & $82.05\pm12.18$ & $84.38\pm05.78$ \tabularnewline
\hline
\multirow{2}{*}{GBM} & <2\,ml (79) & $50.75\pm32.75$ & $65.28\pm23.17$ & $80.42\pm17.11$ & $75.63\pm23.05$ & $90.08\pm06.75$ & $70.80\pm16.30$ & $62.71\pm23.34$ & $86.41\pm09.03$\tabularnewline
& $\geq$2\,ml (2046) & $86.57\pm13.74$ & $87.29\pm11.28$ & $97.34\pm01.32$ & $99.17\pm00.95$ & $95.60\pm02.02$ & $88.96\pm05.04$ & $86.07\pm07.81$ & $92.30\pm03.14$ \tabularnewline
\end{tabular}
}
\label{tab:categorical-seg-results}
\end{table}

\subsection*{Runtime experiments}
A comparison in runtime processing speed using Raidionics for generating the segmentation masks and standardized reports is provided in Table~\ref{tab:runtime-results}. For each CNS tumor subtype, a representative MR scan, with dimension in voxels indicated in parenthesis in the table, was processed five times in a row and speed results were averaged. Two different machines were used: a high-end desktop computer with an Intel Xeon W-2135 CPU (@3.7GHz) and 64GB of RAM (noted Desktop), and a mid-end laptop computer with an Intel Core Processor (i7@1.9GHz) and 16GB of RAM (noted Laptop).

\begin{table}[!b]
\centering
\caption{Segmentation and standardized reporting (RADS) runtime for an average MR scan of each tumor subtype, on two different machines noted Desktop and Laptop. The runtime unit used is indicated in brackets, and MR scan dimensions (voxels) are reported in parenthesis.}
\scalebox{0.80}{
\begin{tabular}{ll|cc|cc|c}
\multirow{2}{*}{CNS type} & \multirow{2}{*}{Machine} & \multicolumn{2}{c|}{Segmentation [s]} & \multicolumn{2}{c|}{RADS [min]} & \multirow{2}{*}{Total [min]}\tabularnewline
& & Brain & Tumor & Registration & Computation & \tabularnewline
\hline
Meningioma & Desktop & $09.16\pm0.21$ & $44.80\pm0.51$ & $02.49\pm0.019$ & $2.15\pm0.019$ & $05.55\pm0.015$\tabularnewline
($256 \times 256 \times 170$) & Laptop & $15.93\pm0.22$ & $112.3\pm5.51$ & $04.92\pm0.127$ & $4.02\pm0.070$ & $11.09\pm0.025$ \tabularnewline
\hline
Metastasis & Desktop & $39.99\pm0.63$ & $83.38\pm0.82$ & $10.56\pm0.059$ & $1.89\pm0.051$ & $14.61\pm0.042$ \tabularnewline
($512 \times 512 \times 513$) & Laptop & $62.51\pm4.21$ & $210.7\pm20.6$ & $21.69\pm0.089$ & $3.52\pm0.124$ & $29.94\pm0.822$ \tabularnewline
\hline
LGG & Desktop & $08.06\pm0.12$ & $45.29\pm0.35$ & $02.44\pm0.015$ & $1.32\pm0.032$ & $04.65\pm0.049$\tabularnewline
($394 \times 394 \times 80$) & Laptop & $16.96\pm3.55$ & $131.7\pm14.3$ & $04.74\pm0.063$ & $2.31\pm0.022$ & $09.53\pm0.316$\tabularnewline
\hline
GBM Preop. & Desktop & $15.81\pm0.06$ & $53.59\pm0.69$ & $03.31\pm0.018$ & $1.66\pm0.020$ & $06.17\pm0.054$ \tabularnewline
($320 \times 320 \times 220$) & Laptop & $23.10\pm3.69$ & $138.8\pm16.0$ & $06.44\pm0.077$ & $3.11\pm0.111$ & $12.27\pm0.554$ \tabularnewline
\hline
GBM Postop. & Desktop & $41.27\pm0.18$ & $96.44\pm0.70$ & $02.10\pm0.072$ & $0.01\pm00.00$ & $04.42\pm0.077$ \tabularnewline
($256 \times 256 \times 176$) & Laptop & $71.23\pm4.06$ & $170.2\pm16.1$ & $06.46\pm0.334$ & $0.08\pm0.006$ & $10.53\pm0.423$ \tabularnewline
\end{tabular}
}
\label{tab:runtime-results}
\end{table}

For preoperative tasks, an average of one minute is necessary for generating the tumor segmentation mask, and around six minutes in total for computing the standardized report, using the desktop machine. When using a computer with less computational power, the processing speed is halved on average as indicated by the laptop runtime results.
A large runtime variation can also be noticed from the MR scan dimensions. The image registration to MNI space takes three times longer over the highest resolution images, whereas the computation of the standardized report in itself remains around two minutes overall. Considerable speed improvement would be obtained by downsampling the high resolution MR scans before computing the standardized report, especially when processing a patient cohort.
Regarding the postoperative task, a combination of MR scans is required, including T1-weighted and contrast-enhanced T1-weighted sequences. As such, the brain must be segmented independently in four MR scans, increasing the runtime to 40 seconds on average. Similarly, the tumor must be segmented in both pre- and postoperative contrast-enhanced T1-weighted scans. Hence, the generation of a postoperative standardized report requires from five minutes on the desktop machine to ten minutes on the laptop.
In general, brain segmentation is three to four times faster to perform than tumor segmentation due to a design choice. On one hand, the brain segmentation model is run in single-shot inference over a downsampled version of the whole input MR scan. On the other hand, iterative inference is performed over patches from the input MR scan with tumor segmentation models. As a result, the segmentation runtime increases with the number of patches to process.

\section*{Discussion}
In this study, the Raidionics software has been presented for enabling the use of CNS tumor segmentation models and standardized reporting methods, through a carefully designed GUI. The software is the first to provide access to competitive preoperative segmentation models for the most common CNS tumor types (i.e., GBM, LGG, meningioma, metastasis) in addition to an early postoperative GBM segmentation model. Standardized reports can be generated to automatically and reproducibly characterize a preoperative tumor or provide a postsurgical assessment through volume and extent of resection computation. Furthermore, new preoperative CNS tumor segmentation models were trained using the AGU-Net architecture, and thoroughly validated. The use of a patch-wise approach, conversely to the full volume approach, allows for a more efficient segmentation of smaller tumors, with a drop in performances noticed below a 2\,ml volume cut-off.

Previously, the preoperative CNS segmentation models included in the Raidionics prototype were all trained following a full volume approach~\cite{bouget2022preoperative}, with a well-identified drawback in the inability to segment accurately the smallest structures. The use of patch-wise techniques leads to improved recall performances, but sometimes at the expense of precision due to the generation of more false positive predictions. By training our AGU-Net architecture with a patch-wise approach, higher recall performances were obtained whereas satisfactory precision performances were retained. For preoperative glioblastoma segmentation, an average F1-score of almost 97\% is being reported in this study, higher than the 94\% reported over the same patient cohort using the nnU-Net architecture~\cite{bouget2021glioblastoma}. Models trained with both architectures reached 98\% recall, but the AGU-Net discriminate better with up to 95\% precision.
Whereas overall segmentation performances are satisfying, the performances for the smallest tumors with a $<2$\,ml volume still needs to be improved, especially for early-stage tumor detection during screening or incidental finding.

Currently, the standardized preoperative reports provide a CNS tumor analysis focusing heavily on overall location in the brain and respective location in relation to cortical and subcortical structures. As it stands, the set of extracted characteristics may not be sufficiently exhaustive to be used as part of preoperative surgical assessment meetings. Similarly, the postoperative standardized report limits itself to the most important parameter to assess, the extent of resection. Nevertheless, the robust, reproducible, and standardized computation of such reports is the first of its kind to be freely available and the list of computed characteristics can be extended in the future.
As a side note, Raidionics allows the user to provide already acquired tumor segmentation masks (i.e., manually or semi-automatically) for computing the standardized report. Bypassing the automatic segmentation process can be extremely valuable as segmentation models are not perfect and might fail to segment, either fully or partially.

Compared to the initial prototype, Raidionics is now a well-rounded and stable software solution, working across all major operating systems, with a welcoming GUI. Clinical end-users can generate the needed segmentations or reports in a few clicks, over single cases or patient cohorts, and directly visualize the results within the software. Users with programming knowledge have the possibility to circumvent the use of the GUI altogether. A stand-alone backend library, used for running the segmentation and standardized reporting tasks, has also been made available both as a Python package and as a Docker image. Furthermore, the possibility is given to integrate the Raidionics backend into other frameworks with relative ease. For example, a plugin for 3D Slicer has been developed, using the Docker image for running all computation. In a similar fashion, a direct integration towards PACS or inside OHIF can be established in the near future.
To include a larger assortment of segmentation models in the future, the decision was made to use an open standard for machine learning interoperability (i.e., ONNX). Models trained using the most common deep learning frameworks (i.e., TensorFlow, PyTorch) can be easily converted to ONNX and deployed inside Raidionics. In addition, the ONNX runtime libraries have been designed to maximize performance across hardware and should provide a better user experience.

The Raidionics environment is under active development with the intent to release more segmentation models and expand the list of characteristics constituting the standardized reports. Better postoperative segmentation models will be investigated, not only for remnant tumor detection but also postoperative complications like hemorrhages or infarctions. For research and benchmarking purposes, a metrics computation module and heatmap location generator module are prospective components to be included. Finally, open-source and state-of-the-art models could be included to extend the record of brain structures to segment.
All users are invited to provide feedback for improvement or contribute code directly to the Raidionics environment at~\url{https://github.com/raidionics}. In addition, project collaborations for testing the software in clinical practice or data-sharing for the training of better models are more than welcome. 

\section*{Data Availability Statement}
The data analyzed in this study is subject to the following licenses/restrictions: patient data are protected under GDPR and cannot be publicly distributed. Requests to access these datasets should be directed to David Bouget (david.bouget@sintef.no) for consideration.

\bibliography{sample}



\section*{Acknowledgements}
All clinical practitioners from the different hospitals who contributed by providing MR scans and manual tumor annotations.
Data were processed in digital labs at HUNT Cloud, Norwegian University of Science and Technology, Trondheim, Norway.

R.H.H. is supported by a grant from The Research Council of Norway, grant number 323339. D.B., I.R., and O.S. are partly funded by the Norwegian National Research Center for Minimally Invasive and Image-Guided Diagnostics and Therapy.

\section*{Author contributions statement}
D.B. and R.H.H. conceived and conducted the experiments; D.A. and V.G. designed the software; A.P. and D.B. developed the software; D.B. analyzed the results; I.R. and O.S. acquired the fundings; All authors read and approved the final manuscript.

\section*{Additional information}

\textbf{Accession codes}
The Raidionics environment with all related information is available at~\url{https://github.com/raidionics}.
More specifically, all trained models can be accessed at~\url{https://github.com/raidionics/Raidionics-models/releases/tag/1.2.0}, the Raidionics software can be found at~\url{https://github.com/raidionics/Raidionics}, and the corresponding 3D Slicer plugin at~\url{https://github.com/raidionics/Raidionics-Slicer}.

\textbf{Competing interests} The authors declare no conflict of interest. The funders had no role in the design of the study; in the collection, analyses, or interpretation of data; in the writing of the manuscript; nor in the decision to publish the results.



\end{document}